**PRE-PRINT**

# Time Dependent Dielectric Breakdown in 4H-SiC power MOSFETs under positive and negative gate-bias and gate-current stresses at 200°C

P. Fiorenza[1], F. Cordiano[2], S. M. Alessandrino[2], A. Russo[2], E. Zanetti[2], M. Saggio[2], C. Bongiorno[1], F. Giannazzo[1], F. Roccaforte[1]

[1] CNR-IMM, Strada VIII n. 5 – Zona Industrale, 95121, Catania, Italy
[2] STMicroelectronics, Stradale Primosole 50, 95121, Catania, Italy
*patrick.fiorenza@imm.cnr.it

**Abstract**

The gate oxide lifetime in 4H-SiC power MOSFETs is typically assessed at fixed and constant gate bias stress, monitoring the time-dependent dielectric breakdown (TDDB). In this work, the TDDB of 4H-SiC power MOSFETs was measured at wafer-level considering three different insulator electric fields stress conditions– either for positive and negative gate bias values. In particular, the TDDB was measured at 200 °C under the following conditions: (i) 3 (positive or negative) fixed gate bias values; (ii) 3 (positive or negative) fixed gate current values; (iii) 3 different insulator fields varying the gate bias stress voltage in each device after performing C-V measurements to determine the actual gate insulator thickness. The three methods provided different lifetime predictions at low oxide field (under standard device operation). The physical explanation of these findings can be found taking into account the device design across the source-body-JFET junction and the transient trapping phenomena at the $SiO_2$/4H-SiC interface.

**Introduction**

The increasing demand of 4H-SiC power MOSFETs for automotive applications imposes the introduction of more severe reliability criteria [1]. Hence, the scientific community is establishing new characterization protocols suitable for wide-bandgap semiconductor materials [2,3]. In particular, the time-dependent dielectric breakdown (TDDB) is used with some approximations (i.e. E-model) to predict the failure rate of the device under standard operative conditions. However, in 4H-SiC MOSFETs some physical features must be taken into account to assess the reliability and correctly predict the device lifetime. Hence, the characterization methods used for silicon need to me appropriately modified for SiC devices. In this paper, the TDDB measurements were carried out at 200 °C at wafer-level. The following conditions were used: (i) 3 (positive or negative) fixed gate bias ($V_G$) values (CVS); (ii) 3 (positive or negative) fixed gate current values (CCS); (iii) 3 different oxide fields ($E_{ox}$) where the gate bias (not fixed) for each device is chosen after a capacitance-voltage (C-V) measurement to determine the actual gate insulator thickness ($t_{ox}$) and using the approximation $E_{ox}=V_G/t_{ox}$ (CFS). By comparing the different methods allowed to separate the intrinsic insulator breakdown kinetics, the source-body-JFET morphology, and the wafer-level processing variability.

**Experimental**

Vertical power MOSFETs were fabricated on n-type 4H-SiC epi-layers ($N_D$ 1×10$^{16}$ cm$^{-3}$), with a p-type Al-implanted body region ($N_A$~10$^{17}$ cm$^{-3}$). The gate oxide was a ~50 nm thick deposited $SiO_2$ layer [4]. Post oxide deposition annealing is performed at 1150 °C in NO [5]. The electrical propertied of the MOSFETs were obtained using a MPI probe station, using a Keysight B1505A parameter analyser. The MOSFETs nanoscale structure was investigated by Transmission Electron Microscopy (TEM) analyses with a JEOL ARM200CF microscope.

**Results and Discussion**

Figs. 1a and 1b show the CVS gate current transients ($I_G$-t) of the vertical 4H-SiC MOSFETs at three different fixed positive and negative gate bias ($V_G$) values respectively. As can be seen in Fig. 1a the different positive $V_G$ values produced a different initial $I_G$ value that varied with time. The current first increased, and then decreased with time, until the breakdown (BD) is reached. On the other hand, Fig. 1b shows the different negative $V_G$ values producing a similar initial $I_G$ value (at t=0s) that is decreased with time till BD. Figs. 1c and 1d show the Weibull plots obtained from Figs. 1a and 1b. It can be noticed that all the distributions possess a unique slope. Hence, no random event have been found. Data depicted in Fig. 1b shows a gate current that is not sensitive to the $V_G$ variation.

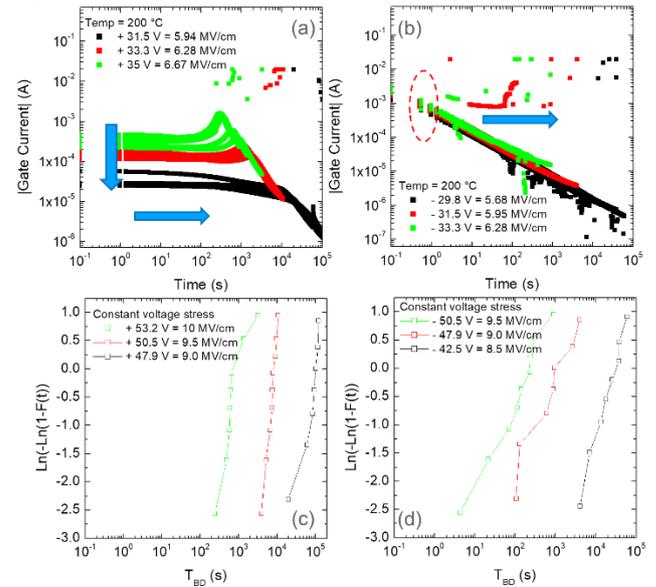

**Fig. 1.** CVS: $I_G$ as a function of time for different positive (a) and negative $V_G$ (b). Their TDDB Weibull plots are also reported in (c) and (d).

This latter can be understood looking at the negative part of the $I_G$-$V_G$ shown in Fig.2, where the first and the second negative bias sweeps resulted in a different gate current flow. In fact, it has been already demonstrated that transient trapping phenomena may occur by an abrupt $V_G$ biasing [6], resulting in a variable oxide field application at fixed gate bias stress, which also affect the gate current. This suggested us to use the $I_G$ as controller of the real oxide field value. Fig.2 top x-axis shows also the simulated oxide field that fits the steady gate current flow. Figs. 3a and 3b show the gate bias transients ($V_G$-t) of the vertical 4H-SiC MOSFETs at three different fixed negative and positive gate current ($I_G$) values respectively (CCS). This procedure allowed the fine control of the oxide field resulting in very steep Weibull plots depicted in Fig. 3c and 3d, respectively. These results indicated the intrinsic insulator BD kinetics.

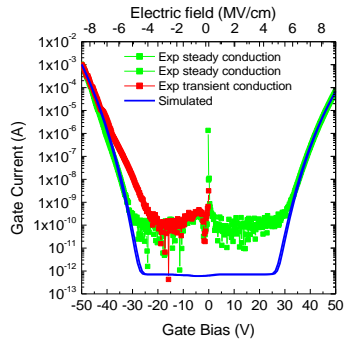

**Fig. 2.** Repeated $I_G$-$V_G$ curves on the MOSFETs compared with the TCAD simulated one both as function of $V_G$ and oxide electric field.

Due to the variability of the physical properties of the devices across the 150 mm wafer, a third method is also used. It is similar to the first described method, but the gate bias stress is not fixed and it is chosen for each fresh device after a C-V measurement used to determine the oxide thickness and converting the gate bias in oxide field by the approximation $E_{ox}=V_G/t_{ox}$. The results of the CFS are shown in Figs. 4a and 4b for three different positive and negative oxide fields. The approximation $E_{ox}=V_G/t_{ox}$ suffers on the lack of control of the local flat band voltage variation and on the local oxide field distribution.

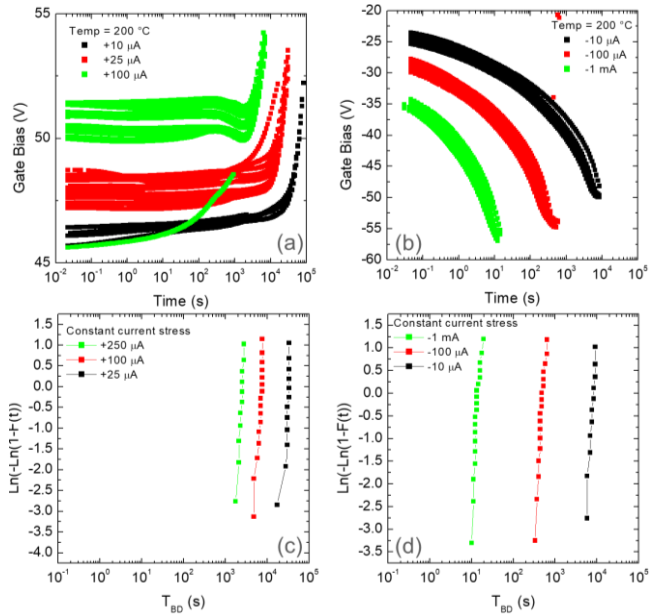

**Fig. 3.** CCS: $V_G$ as a function of time for different positive (a) and Negative $I_G$ (b). Their TDDB Weibull plots are also reported in (c) and (d).

In fact, Fig. 5a and 5b show the TCAD simulation of the electric field in the 4H-MOSFET at $V_G$=+10V and $V_G$=-10V, respectively. Fig 5c shows the comparison of a cutline across the unitary cell in the oxide. As can be noticed, positive $V_G$ produced a nearly constant oxide field, while $V_G$ resulted in a much larger oxide field in the source region compared to the JFET. This result emphasize the concept that constant bias stress may suffer of a lack of control of the actual electric field in the gate oxide layer across the unitary cell of the power MOSFET.

A cross-section TEM micrograph (Fig.6a) in the source-body-JFET region demonstrated the presence of some steps and local oxide thickness variation. All these structural features may enhance the oxide field, accelerating its degradation and lowering the device reliability.

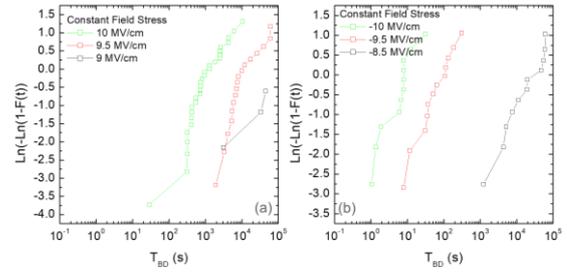

**Fig. 4.** CFS: $V_G$ as a function of time- chosen after the $t_{ox}$ electrical evaluation for different positive (a) and Negative $I_G$ (b). Their TDDB Weibull plots are also reported in (c) and (d).

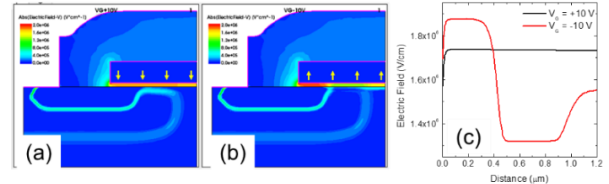

**Fig. 5.** TCAD simulations of the electric field for $V_G$ =+ 10V (a) and for $V_G$ =- 10V (b). (c) The cutline of the oxide field across the source-body-JFET region is shown.

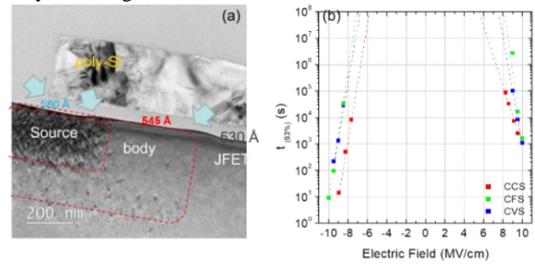

**Fig. 6.** (a)TEM cross section in the urce-body-JFET region (b) $t_{63\%}$ as function of the oxide electric field extracted for the three methods; CCS, CFS, CVS.

Fig. 6b shows the comparison of the $t_{63\%}$ extracted from the Weibull plots for different oxide field methods. As can be noticed, the constant current stress points are located in the internal oxide field region, thus demonstrating that constant voltage stress methods are optimistic on the TDDB prediction at low fields.

**Conclusions**

The TDDB of 4H-SiC power MOSFETs was measured at 200 °C comparing three different methods. The constant voltage stress (CVS) methods resulted in an optimistic lifetime prediction, while constant gate current method (CCS) resulted in a different field device lifetime predictions. This difference has been explained by the occurrence of transient trapping phenomena, and by specific morphological features at the source-body-JFET interface that may enhance the oxide field, reducing the device reliability.

**Acknowledgments**

This paper has been partially supported by Horizon EU Advances in Cost-Effective HV SiC Power Devices for Europe's Medium Voltage Grids (AdvanSiC). The AdvanSiC project has received funding from the European Union's Horizon Europe programme under grant agreement No 101075709. Views and opinions expressed are however those of the author(s) only and do not necessarily reflect those of the European Union. Neither the European Union nor the granting authority can be held responsible for them.